\documentclass[]{JHEP3}
\usepackage{epsfig,amsmath,amssymb}

\def\nn{{\nonumber}}
\newcommand{\bea}{\begin{eqnarray}}
\newcommand{\ena}{\end{eqnarray}}

\title{Hydrodynamics with conserved current from the gravity dual}
\author{Jin Hur\\School of Computational Sciences, Korea Institute for Advanced Study, Seoul 130-722, Korea\\E-mail: \email{hurjin@kias.re.kr}}
\author{Kyung Kiu Kim\\School of Physics, Korea Institute for Advanced Study, Seoul 130-722, Korea\\E-mail: \email{kimeagle@kias.re.kr}}
\author{Sang-Jin Sin\\Department of Physics, Hanyang University, Seoul 133-791, Korea\\E-mail: \email{sjsin@hanyang.ac.kr}}

\abstract{We determine the structure of the hydrodynamics with conserved current,
using the gauge/gravity duality of charged black-hole background.
It turns out that even in the presence of the external electromagnetic field at the boundary, bulk Einstein equation is equivalent to the boundary  conservation of energy momentum tensor and that of current.
As a consequence, the thermal conductivity and electric conductivity
are calculated in terms of the parameters of the fundamental theory.
We find that Wiedermann-Franz law hold with Lorentz number $1/e^2$.}

\begin{document}

\section{Introduction}
After the discovery of consistency of AdS/CFT~\cite{ads/cft,gkp,w}
  and that of RHIC experiment on the viscosity/entropy-density
ratio~\cite{pss0,kss,bl},
much attention has been drawn to the calculational scheme provided
by string theory. Some attempt has been made to map the entire process of RHIC experiment in terms of the gravity dual~\cite{ssz}.
The way to include chemical potential in the theory was
figured out in~\cite{ksz,ht} and phases of these theories were also discussed in D3/D7 setup \cite{nssy1,kmmmt,nssy2,ht}. The essential feature is that the global U(1) charges in the boundary corresponds to the local U(1) charges in the bulk.

The gauge/gravity duality holds for the theory with $N=4$ Super Yang Mill system which is far from the real QCD. Therefore we need to search quantities which does not depends on the details of the theory like  $\eta/s$.
The hydrodynamics approach is especially useful in this respect
since it is about dynamics of long wavelength and low frequency limit. Also, due to high temperature nature, all the fermions are decoupled.

Recently, in a very interesting paper \cite{minwalla01},  the authors  showed that the hydrodynamic equation  in the boundary theory is equivalent to the Einstein equation in the bulk.
Usually the dissipative parts of the current and energy momentum tensor are constructed from the second law by considering the entropy current \cite{landau}. The method of reference \cite{minwalla01} has the merit to determine the dissipative part of the energy momentum tensor  without using the second law.
This merit was applied to other different black holes \cite{raamsdonk,forced fluid,Haack,Bhatta}.
  The purpose of this work is to determine the dissipative part of the
conserved current along this line.  We will also get transport coefficients in terms of the data of the Yang-Mill
theory.
Since the dual of the particle number can be regarded as the local charge of the Maxwell field in the bulk, we are lead to consider the charged black holes in AdS space. We consider only up to the first order in the derivative expansion. It turns out that  our expression for the current is precisely agree with
that of Landau-Lifshitz \cite{landau,sonRcharge}. The only difference
is that the hydrodynamics transport coefficients are now determined from the fundamental theory.

The rest of the paper goes as follows:
in  section 2, we give an outline of calculational  procedure following
\cite{minwalla01}.
In section 3, we calculate the structure of conserved current
using the perturbative solution to the first order in derivative expansion, which is enough to determine both thermal and electric conductivity.
In section 4, we give a summary and discussion. In appendix, we list the important formula.

\section{General construction}

\subsection{General procedure of calculation}
 In this section, we explain the general procedure of constructing solution. Our starting point is the Einstein Maxwell theory with a negative cosmological constant. The action is given by
\begin{eqnarray}\label{action}
S = \frac{1}{16\pi G} \int_\mathcal{M} d^5 x \sqrt{-g}~\left[ R + \frac{12}{l^2} \right] &+ & \frac{1}{8\pi G}\int_{\partial \mathcal{M}} d^4 x \sqrt{-\gamma}\Theta -  \frac{1}{4g^2}\int_{\cal M} d^5 x \sqrt{-g} F^2 \nonumber\\
& +& \text{counter terms}~~,
\end{eqnarray}
where the first and second terms are usual Einstein-Hilbert action and cosmological constant action, and third term is the Gibbons-Hawking term, where $\Theta$ is trace of extrinsic curvature $\Theta_{\mu\nu}$. Introducing spaces with boundary, this term should be added to the action. In addition, the counter terms are needed to cancel divergences of the action and boundary stress-energy tensor. This is well studied in \cite{kraus}. Taking a unit, where $l=1$ and $16 \pi G = 1 $, the equations of motion are given by
\begin{eqnarray}
&&R_{IJ}- \frac{1}{2}g_{IJ}R -6 g_{IJ} - \frac{1}{2g^2}\left(F_{IK} {F_{J}}^K - \frac{1}{4}g_{IJ} F^2 \right)= 0\\
&&\nabla_J {F^{J}}_{I}=0~~.
\end{eqnarray}
The solution  we are interested in is  the charged uniform black brane solution, which can be written as
\begin{eqnarray}
&&ds^2 = - r^2 f(r)dv^2 + 2 dv dr + r^2(dx_1^2 +dx_2^2 +dx_3^2 )\\\nn
&&f(r)=  1- \frac{2M}{r^4} +\frac{Q^2}{r^6}= \frac{(r^2- r_+^2)(r^2- r_-^2)(r^2+ r_+^2 + r_-^2)}{r^6}\\\nn
&&F = g\frac{2\sqrt 3 Q}{r^3}dv \wedge dr ~~.
\end{eqnarray}
Here we use the Eddington-Finkelstein coordinate $v=t+r_*$ with $dr_*=dr/ {f}$ to make the metric regular at the horizon $r_+$. It is slightly different from time coordinate '$t$'. In the boosted frame,
it can be written as
\begin{eqnarray}\label{rnboost}
&&ds^2 = - r^2 f(r)( u_\mu dx^\mu )^2 - 2 u_\mu dx^\mu dr + r^2P_{\mu \nu} dx^\mu dx^\nu\\\nn
&&F = - g\frac{2\sqrt 3 Q}{r^3} u_\mu dx^\mu \wedge dr~~,~~A =  \left(eA^{\rm ext}_\mu -  \frac{\sqrt 3 gQ}{r^2} u_\mu \right) dx^\mu \\\nn&&
u^0 = \frac{1}{ \sqrt{1 - \beta_i^2} }~~,~~u^i = \frac{\beta_i}{ \sqrt{1 - \beta_i^2}  }\\\nn&&P_{\mu \nu}= \eta_{\mu\nu} + u_\mu u_\nu~~,
\end{eqnarray}
where $M$, $Q$, $A^{\rm ext}_\mu$ and $\beta^i $'s are constants  parametering solutions. It satisfies the equations of motion.

Now we lift above parameters to the functions of $x^\mu$. Then,  (\ref{rnboost}) is not a solution of the equations of motion any more. To find make it a  solution, we have to add  corrections to it. To do this systematically, first we define following tensors
\begin{eqnarray}
&&W_{IJ} = R_{IJ} + 4g_{IJ}+\frac{1}{2g^2}\left(F_{IK}{F^{K}}_J +\frac{1}{6}g_{IJ}F^2\right)\\&&W_I = \nabla_J {F^J}_I~~.
\end{eqnarray}
The equation of motion is given by the vanishing of  $W_{IJ}$ and $W_I$. Taking parameters as functions of $x^\mu$ gives nonzero $W_{IJ}$ and $W_I$.
By construction these are proportional to derivatives of parameter functions. We will call these terms by source terms. To cancel this source terms, we need to add some corrections to metric and gauge fields. We consider the derivative expansion and  do it order by order.
So   obtaining new solution is reduced to finding these correction terms. Due to the redundancy in metric and gauge field, we can take a gauge choice for the correction terms. Our choice for $n$-th correction terms are
\begin{eqnarray}\label{correction}
&&{ds^{(n)}}^2 = \frac{ k_n(r)}{r^2}dv^2 + 2 h_n (r)dv dr + 2 \frac{j^i_n(r)}{r^2}dv dx^i + r^2 \left(\alpha^n_{ij} -\frac{2}{3} h_n(r)\delta_{ij}\right)dx^i dx^j \\\nn
&&A^{(n)} = a^n_v (r) dv + a^n_i (r)dx^i~~.
\end{eqnarray}
For metric parts, this choice is the same as \cite{minwalla01}. Since, for gauge field part, $a_r^n(r)$ does not contribute to field strength, the choice $a_r^n(r)=0$ is trivial.

Our procedure can be described as follows. First, we consider zeroth order solution. By construction,  $S^{(0)}_{IJ}$ and $S^{(0)}_J$ are zero, and the zeroth order correction terms are also zero naturally. For the first order, we take the parameters in zeroth order solution as functions of $x^\mu$ and expand  around $x^\mu=0$ up to first order in derivative\footnote{We assume that $\beta^i(0)=0$}. Putting these new metric and gauge fields into $W_{IJ}$ and $W_I$,  we obtain the source terms $S^{(1)}_{IJ} = - W_{IJ}$ and $S^{(1)}_{I} = -W_I$. To make these source terms canceled, we consider correction terms in (\ref{correction}) for $n=1$, these make effects in $W_{IJ}$ and $W_I$. So $W_{IJ} = (\text{effect from correction}) - S^{(1)}_{IJ}$ and $W_{I} = (\text{effect from correction}) - S^{(1)}_{I}$, from this we can find corrections.   Assuming we know $(n-1)$ th order solution, we can calculate $n$th derivative order source terms and we can obtain $n$th order correction terms in metric and gauge fields. For general $n$th order, in the Maxwell equation, what we have to do is just solving following equations
\begin{eqnarray}\label{Maxwell}
&&W_v = \frac{f(r)}{r}\left\{ r^3 {a^n_v}' (r) + 4\sqrt 3 g Q h^n(r) \right\}'- S^{(n)}_{v}(r)=0\\\nn
&& W_r = - \frac{1}{r^3}\left\{  r^3 {a^n_v}'(r) +4\sqrt 3 g Q h^n(r)  \right\}'-S^{(n)}_r (r)=0\\\nn
&&W_i = \frac{1}{r} \left\{ r^3 f(r) {a^n_i}'(r) - \frac{2 \sqrt 3 g Q }{r^4} j^i_n (r) \right\}' -  S^{(n)}_i (r)=0~~.
\end{eqnarray}\footnote{Where $'$ means derivative of $r$ coordiante.}
For Einstein equations, the equations are
\begin{eqnarray}\label{Integration}\nn
&&W_{vv} = f(r)\left\{ \frac{4  }{\sqrt 3 g } \frac{Q}{r} {a^n_v}'(r) -8 r^2 h_n(r) -r^2 (r^2 f(r))'{h_n}'(r) -\frac{r}{2}\left(\frac{k'_n(r)}{r}\right)'  \right\}-  S^{(n)}_{vv}(r)=0\\\nn
   &&W_{vr} = 8h_n(r) + (r^2 f(r))' h_n'(r)-\frac{4 }{\sqrt 3 g} \frac{Q}{r^3} {a^n_v}'(r) + \frac{1}{2r}\left(\frac{k_n'(r)}{r}\right)' - S^{(n)}_{vr}(r)=0 \\\nn&&W_{vi} = f(r)\left\{ \frac{\sqrt 3 Q}{g r} {a^n_i}'(r) -\frac{r^3}{2}\left( \frac{{j_n^i}' (r)}{r^3} \right)'    \right\}- S^{(n)}_{vi}(r)=0\\\nn
&&W_{rr} = \frac{1}{r^5}  (r^5 h_n'(r))' -  S^{(n)}_{rr}(r)=0\\\nn
&&W_{ri} = -\frac{\sqrt 3 Q}{g r^3} {a^n_i}'(r) + \frac{r}{2}\left( \frac{{j_n^i}' (r)}{r^3} \right)'-  S^{(n)}_{ri}(r)=0\\\nn
&&W_{ij}\delta^{ij} = 24 r^2 h_n(r) + \frac{3}{r}k_n'(r) + \frac{2 \sqrt 3 Q}{gr} {a^n_v}' (r) + \frac{1}{r^7}( r^{11} f(r) h_n'(r) )'- S^{(n)}_{ij}(r) \delta^{ij} =0\\
&&W_{ij} - \frac{1}{3}\delta_{ij}(\delta^{kl}W_{kl})= -\frac{1}{2r} ( r^5 f(r) {\alpha^n_{ij}}'(r)   )' -  S^{(n)}_{ij}(r) + \frac{1}{3}\delta_{ij}(\delta^{kl}S^{(n)}_{kl}(r))=0~~.
\end{eqnarray}
After solving above equations $n$th derivative order solution can be found.

There is an important point in this step, actually all equations in (\ref{Maxwell})and (\ref{Integration}) are not independent. So we can find relations between some source terms
\begin{eqnarray}\label{constraint}
 &&W_v + r^2 f(r)W_r =0 ~:~ S^{(n)}_v + r^2 f(r)S^{(n)}_r =0\\\nn
 &&W_{vi} + r^2 f(r) W_{ri} =0 ~:~ S^{(n)}_{vi} + r^2 f(r) S^{(n)}_{ri} = 0\\\nn
 &&W_{vv} + r^2 f(r)W_{vr} =0 ~:~ S^{(n)}_{vv} + r^2 f(r) S^{(n)}_{vr} = 0~~.
 \end{eqnarray}
 Thus our integration procedure is that we integrate  $W_r$,$W_i$,$W_{vr}$,$W_{rr}$,$W_{ri}$,$W_{ri}$ and  $W_{ij}$ equations and consider the constraints (\ref{constraint}). We anticipate fluid dynamics equations $\partial_\mu J^\mu =0 ~~,~~\partial_\mu T^{\mu\nu} = 0$ from the constraints.

 To get more practical guide line to our goal, now we describe way to finding correction terms more explicitly. Considering $W_{rr},W_r,W_{ij} \delta^{ij}$ and traceless part of $W_{ij}$, correction terms can be obtained as follows
\begin{eqnarray}
&&h_n(r) = \int_\infty^r dy \frac{1}{y^5} \int_{ }^y x^5 S^{(n)}_{rr}(x)\\\nn
&&a_v^n(r) = \int_{}^r dy \frac{-4\sqrt 3}{y^3} g Q h_n(y) - \int^r_{} dy \frac{1}{y^3} \int^y dx x^3 S^{(n)}_r (x)\\\nn
&&k_n(r)= \int_0^r dx \frac{x}{3}\left[ S^{(n)}_{ij}(x)\delta^{ij} -24 x^2 h_n(x)-\frac{2\sqrt 3}{gx} Q {a^n_v}'(x)-\frac{1}{x^7}(x^{11}f(x)h_n'(x))' \right]\\\nn
&&\alpha^n_{ij} = -\int^r dy \frac{1}{y^5 f(y)}\int^y dx 2x \left(S^{(n)}_{ij}(x)-\frac{1}{3}\delta_{ij}\delta^{kl}S^{(n)}_{kl} \right)~~.
\end{eqnarray}
From the above equations, we can find corrections scalar and tensor corrections. Using the result, we have to check that $W_{vr} = 0$ equation is consistent. Since the remaining equations $W_i =0$ and $W_{ri} =0$ are coupled each other, it is more difficult to solve these than others. These equations are
\begin{eqnarray}
&& \frac{r}{2} \left(\frac{j_i'(r)}{r^3}\right)'-\frac{\sqrt{3} Q}{g r^3} a_i'(r)=S_{r i}^{(n)}(r), \label{appeq1}\\
&& \left(r^3 f(r) a_i'(r)-\frac{2 \sqrt{3} g Q}{r^4} j_i(r)\right)'=r S_i^{(n)}(r). \label{appeq2}
\end{eqnarray}
For the exact solutions to these equations, see appendix \ref{appendixexact}.

We define the chemical potential as
\begin{eqnarray}\label{chemical potential}
\mu = A_0 (r_+) - A_0 (\infty) ~~.
\end{eqnarray}
 One should notice that this is the chemical potential  that couples with the number density, not the charge density. See appendix \ref{appendixg}.

\subsection{Stress Energy Tensor and Current}

 In this subsection, we describe the prescription to calculate boundary stress-energy tensor and current out of bulk quantity briefly. We will follow prescription in \cite{kraus} for stress energy tensor and calculate the boundary current. In order to calculate these tensors, we need to decompose our metric. Our metric decomposition is a well-known ADM decomposition
\begin{eqnarray}
ds^2 = \gamma_{\mu\nu}(dx^\mu + V^\mu dr)(dx^\nu + V^\nu dr) + N^2 dr^2~~,
\end{eqnarray}
where $V^\mu$ is defined by $\gamma^{\mu\nu}V_\nu$, and the $\gamma^{\mu\nu}$ is the inverse of $\gamma_{\mu\nu}$. Under this decomposition, we can obtain the stress energy tensor from the action (\ref{action}) by variation of boundary metric $\gamma_{\mu\nu}$. The stress energy tensor is given by
\begin{eqnarray}\label{emtensor}
  T_{\mu\nu} \equiv \lim_{r \rightarrow \infty }  r^2 \frac{-2}{\sqrt{-\gamma}} \frac{\delta S_{cl}}{\delta\gamma^{\mu\nu}}   =\lim_{r \rightarrow \infty }  r^2  [  -2 ( \Theta_{\mu\nu} - \Theta \gamma_{\mu\nu} + 3\gamma_{\mu\nu} + \frac{1}{2}G_{\mu\nu} ) ]~~,
\end{eqnarray}
where the last two terms came from counter lagrangian which has been studied in \cite{kraus} and subscript "$cl$" means that we have to impose equation of motion. In our metric decomposition, the expression of extrinsic curvature $\Theta_{\mu\nu}$ is
\begin{eqnarray}
\Theta_{\mu\nu} = \frac{1}{2 N} [ \acute{\gamma_{\mu\nu}} - D_\mu V_\nu - D_\nu V_\mu ]~~.
\end{eqnarray}
Another important physical quantity is the boundary current. The current can be obtained by same method of the stress energy tensor. The current is
\begin{eqnarray}\label{current}
J^\mu &=& \lim_{r \rightarrow \infty} r^4 \frac{1}{\sqrt{-\gamma}} \frac{\delta S_{cl}}{\delta \tilde A_\mu} =  \lim_{r \rightarrow \infty} r^4 \frac{N}{g^2} F^{r \mu} \\\nn
&=& \lim_{r \rightarrow \infty} r^4 \frac{1}{g^2N} [ \gamma^{\mu\lambda}( A'_\lambda  - \partial _\lambda A_r ) - V^\lambda \gamma^{\mu\sigma}(\partial_\lambda A_\sigma - \partial_\sigma A_\lambda)  ]~~,
\end{eqnarray}
where $\tilde A_\mu$ is the gauge field which is projected to the boundary.

\section{Perturbative solution and fluid dynamics}

\subsection{Perturbative solution}

 As we discussed in previous section, to obtain higher order derivative solution, we should take the parameters of the solution as functions of boundary coordinates. Let us rewrite the zeroth order solution with functions. The zeroth order solution is
\begin{eqnarray}
g^{(0)}_{IJ} dx^I dx^J &=&-r^2 f(r) u_{\mu } u_{\nu } dx^{\mu } dx^{\nu }-2 u_{\mu } dx^{\mu } dr+r^2 P_{\mu  \nu } dx^{\mu } dx^{\nu }\\\nn A^{(0)}
&=& \left(eA^{\rm ext}_\mu (x) -\frac{\sqrt{3}gQ(x)}{r^2} u_{\mu }(x) \right)dx^{\mu }~~.
\end{eqnarray}
This solution satisfies the Einstein equation and the Maxwell equation in zeroth order.
Notice that we need to distinguish the R-charge coupling $g$ and the external electric charge coupling $e$.
Using the gauge field part and the definition of chemical potential (\ref{chemical potential}), we can read chemical potential as
\begin{eqnarray}
\mu (x) = \frac{\sqrt 3 g Q(x)}{ r_+^2 (x)}~~.
\end{eqnarray}
Putting this zeroth order solution into the general formula of boundary tensors, (\ref{emtensor}) and (\ref{current}), we can get the zeroth order boundary (particle number) current and energy momentum tensor,
\begin{eqnarray}
&&J_{(0)}^{\mu} = \frac{2\sqrt 3 Q}{g } u^\mu :=nu^\mu \\\nn
&&T_{(0)}^{\mu\nu} = 2M (\eta^{\mu\nu} + 4 u^\mu u^\nu)~~.
\end{eqnarray}
These are well-known perfect fluid current and energy momentum tensor.

Next step is to obtain first order corrections. To construct first order correction, we regard parameter $M$,$Q$,$A^{\rm ext}_\mu$ and $\beta_i$ as functions of $x^\mu$ in (\ref{rnboost}) and expand these parameter functions around $x^\mu =0$, then this expansion gives source terms  $S^{(1)}_{IJ}(r,x)$ and $S^{(1)}_I(r,x)$. After a little working, we can find first order source terms. The source terms $S^{(1)}_{I}$ and $S^{(1)}_{IJ}$ are given as follows
\begin{eqnarray}
&&S_{v v}^{(1)}(r)=-\frac{1}{2} \left(r^2 f(r)\right)' \partial _i\beta _i-\frac{3}{r^3} \partial _vM+\frac{3 Q}{r^5} \partial _vQ   \\\nn&&S_{v r}^{(1)}(r)=\frac{\partial _i\beta _i}{r}\\\nn&&S_{v i}^{(1)}(r)=\left(\frac{3 r}{2}+\frac{M}{r^3}+\frac{3 Q^2}{2 r^5}\right) \partial _v\beta _i+\frac{\partial _iM}{r^3}-\frac{\sqrt{3} Q}{r^3} \frac{e}{g} F^{\rm ext}_{vi}\\\nn&&S_{r r}^{(1)}(r)=0\\\nn&&S_{r i}^{(1)}(r)=-\frac{3}{2 r} \partial _v\beta _i\\\nn&&S_{i j}^{(1)}(r)=r \left\{\delta _{i j} \partial _k\beta _k+\frac{3}{2} \left(\partial _i\beta _j+\partial _j\beta _i\right)\right\}
\end{eqnarray}
\begin{eqnarray}
&&S_v^{(1)}(r)=g\frac{2 \sqrt{3}}{r^3} \left(\partial _vQ+Q \partial _i\beta _i\right)\\\nn&&S_r^{(1)}(r)=0\\\nn&&S_i^{(1)}(r)=g \left(-\frac{\sqrt{3}}{r^3} \left(\partial _iQ+Q \partial _v\beta _i\right)-\frac{1}{r} \frac{e}{g}F^{\rm ext}_{vi} \right)~~,
\end{eqnarray}
where $F^{\rm ext}_{vi}\equiv \partial_v A^{\rm ext}_i-\partial_i A^{\rm ext}_v$ is the external field strength tensor.
Using above result, one can easily calculate parts of correction terms,
\begin{eqnarray}
&&h^{(1)} (r) = 0~,~a_v^{(1)} = 0~,~ k^{(1)}(r) = \frac{2}{3}r^3 \partial_i\beta^i\\\nn &&\alpha^{(n)}_{ij} =  \alpha(r)\left\{ (\partial_i \beta_j + \partial_j \beta_i )-\frac{2}{3} \delta_{ij}\partial_k \beta^k \right\}~~,
\end{eqnarray}
where definition of $\alpha(r)$ and it's asymptotic expression are given by
\begin{eqnarray}
\alpha(r)= 3 \int_\infty^{ r } dt  \frac{1}{t^5 f(t)} \int_{r_+}^t ds  s^2 \approx \frac{1}{r}-\frac{r_+^3}{4 r^4}+\frac{2 M}{5 r^5}-\frac{Q^2}{7 r^7}-\frac{M r_+^3}{4 r^8}+O\left(\frac{1}{r}\right)^9~.
\end{eqnarray}
There is another constraint $W_{vr} = 0$ which must be checked by the above information, and it is very easy to check that the above correction terms satisfy the constraint.

Though we can write explicit expressions of $j_i(r)$ and $a_i(r)$ from appendix {\ref{appendixexact}}, it is too long to write in the main text and we only need the asymptotic behaviors of those solutions. In order to write covariant form, we decompose first order vector terms as
\begin{eqnarray}
a_i(r) &=& a_\beta(r) \partial_v \beta_i + a_Q(r) (\partial_i Q+Q\partial_v \beta_i) +a_F(r) F^{\rm ext}_{vi}, \\
j_i(r) &=& j_\beta(r) \partial_v \beta_i + j_Q(r) (\partial_i Q+Q\partial_v \beta_i) +j_F(r) F^{\rm ext}_{vi}.
\end{eqnarray}
With this decomposition, the gauge field can be written as
\begin{eqnarray}
A^{(1)}_\mu(r)=a_\beta(r) u^\nu \partial_\nu u_\mu +a_Q(r) u^\nu F^{(Q)}_{\nu\mu} +a_F(r)u^\nu F^{\rm ext}_{\nu\mu},
\end{eqnarray}
where $F_{\lambda  \nu }^{(Q)}\equiv \partial _{\lambda }\left(Q u_{\nu }\right)-\partial _{\nu }\left(Q u_{\lambda }\right)$. The form of metric is
\begin{eqnarray}
ds^2&=&-r^2 f(r) u_{\mu } u_{\nu } dx^{\mu } dx^{\nu }-2 u_{\mu } dx^{\mu } dr+r^2 P_{\mu  \nu } dx^{\mu } dx^{\nu }+\Bigg[\frac{2 r}{3} u_{\mu } u_{\nu } \partial _{\lambda }u^{\lambda }\\\nn& & - \frac{2}{r^2} \left\{
\frac{1}{2}j_{\beta } (r) u^{\lambda } \partial _{\lambda }\left(u_{\mu } u_{\nu }\right)+  j_Q (r)  u_{\mu } u^{\lambda } F _{\lambda  \nu }^{(Q)}+  j_{F}(r)  u^{\lambda } F^{\rm ext}_{\lambda  \nu } \right\}\\\nn && +2 r^2 \alpha (r) \sigma _{\mu  \nu }\Bigg] dx^{\mu } dx^{\nu }~~,
\end{eqnarray}
where we took the covariant expression and the definitions of $\sigma_{\mu \nu}$ is
\begin{eqnarray}
\sigma ^{\mu  \nu }\equiv \frac{1}{2} P^{\mu  \alpha } P^{\nu  \beta } \left(\partial _{\alpha }u_{\beta }+\partial _{\beta }u_{\alpha }\right)-\frac{1}{3} P^{\mu  \nu } \partial _{\alpha }u^{\alpha }~~.
\end{eqnarray}

The chemical potential in 1st order calculation is
\begin{eqnarray}
\mu =  \frac{\sqrt 3 gQ(x)}{ \left( r_+ + \delta r_+^{(1)}  \right)^2},
\end{eqnarray}
where $\delta r_+^{(1)}$ is the first order correction of the horizon. As a final step for completing our consideration, we have to find $\delta r_+^{(1)}$, so we need to examine the first order correction of  horizon. In general, the horizon is defined by following equation
\begin{eqnarray}
g^{IJ}\partial_I (r -r_H) \partial_J (r-r_H) = 0~~.
\end{eqnarray}
 In zeroth order, we already know the horizon $r_H = r_+$. As a check, zeroth order equation is just $g^{rr}=0$, then it's solution is the right value $r_+$. In the first order, the equation is
  \bea g^{rr} -2 g^{r\mu} \partial_\mu r_H =0 ,\ena
   but there is no correction to the solution and $\delta r_+^{(1)}$ vanishes. So the chemical potential keeps same expression,
 \begin{eqnarray}
 \mu = 
 \frac{\sqrt 3 gQ(x)}{  r_+ ^2 (x)}~~,
 \end{eqnarray}
 where we have to notice that $Q$ and $r_+$ are not constants.

\subsection{Boundary tensors and fluid dynamics}
~~~~~Putting our source terms into the equations (\ref{constraint}), they give constraints
\begin{eqnarray}
&&3 \partial _vM+4 M \partial _i\beta _i=0  \\\nn &&\partial _iM+4 M \partial _v\beta _i=\sqrt{3} Q   \frac{e}{g}F^{\rm ext}_{vi}\\\nn &&\partial _vQ+Q \partial _i\beta _i=0~~.
\end{eqnarray}
we can rewrite these in covariant form,
\begin{eqnarray}\label{conservation}
&&\partial_\mu T_{(0)}^{\mu\nu} = 2\sqrt 3  Q \frac{e}{g}F_{\rm ext}^{ \mu\nu}u_\mu \\ \nn&&
\partial_\mu J_{(0)}^\mu = 0~~.
\end{eqnarray}
Since there is a derivative, these are exact first order (non)conservation equations.
The right hand side is energy and momentum inflow sourced by the external fields.
 When external fields are absent,  we obtain the usual fluid dynamics system with conserved current.
   We expect that the relation
 \begin{eqnarray}
 \partial_\mu T^{\mu\nu}  =  eF_{\rm ext}^{\mu\nu}J_\mu~~,~~\partial_\mu J^\mu = 0~~,
 \end{eqnarray}
 holds to all order.

 \begin{figure}[t]
\centering
\includegraphics{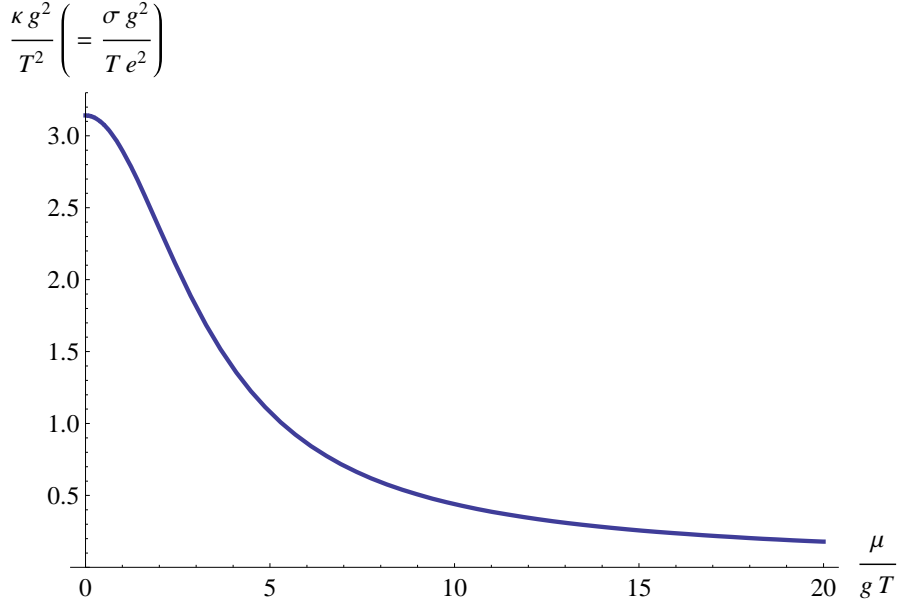}\caption{The coefficient of thermal conductivity or The electrical conductivity}\label{gp1}
\end{figure}

We know the asymptotic forms of metric and gauge fields.
This information is enough to calculate the first order boundary stress-energy tensor and the current. Using our first order metric and equation (\ref{emtensor}), we can find the first order stress-energy tensor
\begin{eqnarray}
T_{\mu\nu} = 2M (\eta_{\mu\nu} + 4 u_\mu u_\nu  ) - 2 r_+^3\sigma_{\mu\nu}~~
:=P(\eta_{\mu\nu} + 4 u_\mu u_\nu  ) - 2 \eta \sigma_{\mu\nu}~~,
\end{eqnarray}
from which we can read off the viscosity $\eta=r_+^3$.
By restoring $1/(16\pi G)=N_c^2/8\pi^2$,  which was set to be 1, we get
\begin{eqnarray}
\eta=\frac{N_c^2}{8\pi^2}r_+^3
=\frac{\pi N_c^2T^3}{8} \left(\frac{1}{2}+\sqrt{\frac{1}{4}+\frac{1}{6}
\left(\frac{\mu}{ g\pi T}\right)^2}\right)^3
.
\end{eqnarray}
where we have used $r_+=\frac{\pi T}{2}\left(1+\sqrt{1+\frac{2}{3}(\mu/g\pi T)^2} ~\right)$.
In addition, we can also calculate the first order boundary current from (\ref{current}). The current is
\begin{eqnarray}
J^\mu &=& J_{(0)}^\mu + J_{(1)}^\mu \\ \label{first order current}
J_{(1)}^{\mu }&=& \frac{1}{g} \left\{-2 \sqrt{3} Q \frac{j_{\beta }\left(r_+\right)}{r_+^4} u^{\lambda } \partial _{\lambda }u^{\mu }+\left(-2 \sqrt{3} Q \frac{j_Q\left(r_+\right)}{r_+^4}-\frac{\sqrt{3}}{r_+}\right) u^{\lambda } F ^{(Q)}{}_{\lambda }{}^{\mu } \right.\nonumber\\
&& \left.+\left(-2 \sqrt{3} Q \frac{j_F\left(r_+\right)}{r_+^4}+\frac{e}{g}r_+\right) u^{\lambda } F ^{\rm ext}{}_{\lambda }{}^{\mu }\right\},
\end{eqnarray}
where the $j_\beta(r_+)$, $j_Q(r_+)$ and $j_F(r_+)$ are values of each function at the horizon, which can be read from (\ref{j at rp}) as follows
\begin{eqnarray}
\frac{j_{\beta }\left(r_+\right)}{r_+^4}=\frac{2 \left(2 r_+^6+Q^2\right)}{8 M r_+^3}~~,~~\frac{j_Q\left(r_+\right)}{r_+^4}=-\frac{Q}{8 M r_+^3}~~,
~~\frac{j_{F}\left(r_+\right)}{r_+^4}=-\frac{e}{g}\frac{\sqrt{3} Q}{8 M r_+}~~
\label{jis}.
\end{eqnarray}
To find physical quantities, we need to change the above equation into familiar form. Using (\ref{conservation}) and the temperature of the black hole, we can obtain the final expression of the current which contains dissipative current and contribution from the external field.
After some algebra as summarized in the appendix \ref{relations}, the expression  turns
out to be simplified and given by
 \begin{eqnarray}\label{familiar current}
 J_{(1)}^\mu  &=& - \frac{\pi ^2 T^3 r_+^7}{4g^2 M^2}  P^{\mu  \nu } \partial _{\nu }\frac{\mu }{T}
+ \frac{\pi ^2 T^2 r_+^7}{4g^2 M^2} u^{\lambda } eF ^{\rm ext}{}_{\lambda }{}^{\mu }~~,\\
 &:=& -\kappa  P^{\mu  \nu } \partial _{\nu }\frac{\mu }{T}+\frac{\sigma}{e} u^{\lambda } F ^{\rm ext}{}_{\lambda }{}^{\mu }~~,
\end{eqnarray}
where $1/e$ in $\sigma/e$ is inserted to read off the electric conductivity from the electric current $eJ$ rather than the number current $J$. See appendix \ref{appendixg}.

Now the coefficient of thermal conductivity $\kappa $  and the electrical conductivity are given by
\begin{eqnarray}
\kappa =  \frac{\pi ^2 T^3 r_+^7}{4 g^2M^2}, \;\;\quad
\sigma =\frac{\pi ^2 e^2T^2 r_+^7}{4g^2 M^2}.
 \end{eqnarray}
To see that the second one is  actually the electric conductivity, we notice that
  $J_{\text{\rm ext}}^{\mu }=\sigma  u^{\lambda } F^{\rm ext}_{\lambda }{}^{\mu }$ becomes in non-relativistic limit  $J_{\text{\rm ext}}=\sigma (E +v\times B)$ with the $\sigma$ given above.
It is easy to show that $\kappa$ can be also written as
\begin{eqnarray}
\kappa={r_+\over 4g^2} \frac{(2-a)^2}{(1+a)^2}, \;\;\text{with } a=Q^2/r_+^6
\end{eqnarray}
so that our result agrees precisely with that of \cite{sinapctp}, which was obtained from the Kubo formula.
Our result of conductivity goes to the known
result in \cite{yaffe,karch} for zero charge case \footnote{The result is different by a constant factor of $\pi$.}.
The   relation between the thermal and
heat conductivity, so called Wiedermann-Franz law:
\begin{equation}
\kappa=\sigma T/e^2,
\end{equation}
is manifest in our case with Lorentz number $1/e^2$.
Considering the complication of the definition of
temperature in the charged black hole and also the difference in the origin of two terms, it is rather surprising to have
such similar and simple recombination such that
$T \sigma$ and $\kappa$ is proportional to each other.
Using
 $   2M=r_+^4(1+\mu^2/(3g^2r_+^2))$ and
 $r_+=\pi T(1+\sqrt{1+2(\mu/g\pi T)^2/3})/2$,
 we can express $\kappa/T^2$ and $\sigma/T$ as a function of $\mu/gT$ only:
 \bea
 \frac{ \kappa}{T^2}= \frac{ \sigma}{Te^2}= \frac{2\pi^2}{g^2} \frac{1 + \sqrt{
  1 +  \frac{2}{3} (\frac{\mu}{g\pi T})^2}}{\left(1-3\sqrt{
  1 +  \frac{2}{3 } (\frac{\mu}{g\pi T})^2}\right)^2}
   \ena
Fig. \ref{gp1} shows this as a function of $ {\mu}/{gT}$.

We can also calculate the thermal conductivity,
\begin{eqnarray}
\kappa_T = \left(  \frac{\epsilon+P}{\rho T} \right)^2 \kappa  =
4\pi^2 \cdot \frac{g^2}{16\pi G}\cdot \frac{\eta T}{\mu^2} .
\end{eqnarray}
Notice that $\frac{1}{16\pi G}=\frac{N_c^2}{8\pi^2}$ and
 $\frac{g^2}{16\pi G}$ is 1 and $N_c/2N_f$ for  R-charge
and Baryon charge respectively \cite{sinRN}.

 \section{Discussion}

 We use gauge/gravity duality to determine the structure of the
 fluid dynamics in the presence of the conserved current and calculated
 thermal conductivity as well as the electrical conductivity in the presence of the the external electric field.
 Since the dual of the particle number can regarded as the local charge of the Maxwell field in the bulk, we used the charged
black holes in AdS space.  For our purpose,
the first order in the derivative expansion.

While the
determination of the current especially the dissipative part is of highly interesting, going second order or higher order is less interesting from the experimental point of view: what RHIC experiment discovered is  that we should neglect the dissipation part almost completely. This perfect liquid behavior
is a hall mark of the RHIC experiment. However,   LHC experiment
may show different behavior due to
its  much higher collision energy scale. Therefore
calculating the transport coefficients may be of some importance for future experiment. It would be also interesting to calculate the similar quantities for other ads space for the application to the solid state physics or M2, M5 brane theories. It would be also interesting to see if this method can help to settle problems of fluid mechanics. \cite{tk}.

\section*{Acknowledgements}
The work of SJS  was supported by
the SRC Program of  the KOSEF through the Center for Quantum
Space-time(CQUeST) of Sogang University with grant number R11 - 2005
- 021 and also by KOSEF Grant R01-2007-000-10214-0.

\section*{Note added}
At the end stage of the this work, we received two papers \cite{Banerjee:2008th,german} whose contents are partly overlapping with present paper.  Therein, external fields are not included  but $T_{\mu\nu}$ and $J_\mu$ are calculated up to second order.

\appendix
\section{Number current v.s charge current: Fixing $g$ factors.} \label{appendixg}
For the lagrangian
\bea
L=-\frac{1}{4g^2}F_{\mu\nu}^2 + J_\mu A^\mu
\ena
  the equation of motion is
\bea
\frac{1}{g^2}\partial^\mu F_{\mu\nu}=J_\nu
\ena
If we rewrite the action in terms of the rescaled variable
$A_\mu/g:= A'_\mu$,
\bea
L=-\frac{1}{4}{F'}_{\mu\nu}^2 + gJ_\mu A'^\mu, \quad
\partial^\mu F'_{\mu\nu}=gJ_\nu:=J'_\nu.
\ena
therefore if $J^0=n$ is the number density, $J'_0=gn$ is the charge density and $J_\mu$ is the number current.
 Also the electromagnetic potential is $A'_{\mu}=A_{\mu}/g$.

The number current density
$J_\mu= 2\sqrt{3}Q u_\mu /g$ so number density is related to the charge parameter $Q$ by
\bea n=2\sqrt{3}Q/g.\ena
The chemical potential for the electric charge density is $\mu'=A'(\infty)-A'(0)=\sqrt{3}Q/r_+^2 $. From the
  $A.J=A'.J'$ or $\mu n= \mu' J'_0$ and  $J'_0=gn$, we get $\mu=g\mu'$.
Then the chemical potential $\mu$
that couples to the number density is $\mu=  A_0(\infty)-A_0(0)$ as was stated in the main text (\ref{chemical potential}).
Similar relation hold for electric charge as well as the R-charge.

 \section{The exact expression for of $j_i(r)$ and $a_i(r)$} \label{appendixexact}
In this appendix we will present exact expressions of $j_i(r)$ and $a_i(r)$ which can be obtained from coupled differential equations (\ref{appeq1}) and (\ref{appeq2}). Integrating (\ref{appeq2}) from $r=r_+$ to $r=\infty$, we get
\begin{eqnarray}
r^3 f(r) a_i'(r)-2 \sqrt{3} g Q \left(\frac{j_i(r)}{r^4}-\frac{j_i\left(r_+\right)}{r_+^4}\right)=\int _{r_+}^rdx x S_i(x), \label{appeq2i}
\end{eqnarray}
where we used the fact that $f(r)$ is zero and $a_i(r)$ is not singular at $r=r_+$. Using (\ref{appeq1}) to eliminate $a_i'(r)$ in (\ref{appeq2i}), we get a second order differential equation of $j_i(r)$
\begin{eqnarray}
j_i{}^{\prime\prime }(r)-\frac{3}{r} j_i'(r)-\frac{12 Q^2}{r^8 f(r)} j_i(r)  = \zeta_i(r), \label{appeqj}\\
\end{eqnarray}
where
\begin{eqnarray}
\zeta_i(r)&\equiv&-\frac{12 Q^2}{r^4 f(r)} \frac{j_i\left(r_+\right)}{r_+^4}+2 r^2 S_{r i}(r)+\frac{2 \sqrt{3} Q}{g r^4 f(r)} \int _{r_+}^rdx x S_i(x) \nonumber\\
&=&-3 r \partial _v\beta _i-\frac{2 \sqrt{3} Q}{r^4 f(r)} \left(2\sqrt3 Q \frac{j_i\left(r_+\right)}{r_+^4}-\sqrt{3} \left(\frac{1}{r}-\frac{1}{r_+}\right) \left(\partial _iQ+Q \partial _v\beta _i\right) \right.\nonumber\\
&& \left.+\left(r-r_+\right) \frac{e}{g}F_{v i}^{\text{\rm ext}}\right).
\end{eqnarray}
Two linearly independent homogeneous solutions of this equation are
\begin{eqnarray}
j_{H_1}(r)&=&r^4 f(r) \\
j_{H_2}(r)&=&j_{H_1}(r) \int _r^{\infty }\frac{x^3 dx}{j_{H_1}(x){}^2}=r^4 f(r) \int _r^{\infty }\frac{dx}{x^5 f(x)^2}.
\end{eqnarray}
A particular solution to (\ref{appeqj}) is found by using method of variation of parameters
\begin{eqnarray}
j_P(r)=b_1(r) j_{H_1}(r)+b_2(r) j_{H_2}(r),
\end{eqnarray}
where
\begin{eqnarray}
b_1(r)&=&-\int _r^{\infty }dx \frac{j_{H_2}(x) \zeta _i(x)}{x^3}, \\
b_2(r)&=&r^3 \partial _v\beta _i+\int _r^{\infty }dx \left(\frac{j_{H_1}(x) \zeta _i(x)}{x^3}+3 x^2 \partial _v\beta _i\right).
\end{eqnarray}
Note that $3 x^2 \partial _v\beta _i$ term cancels the divergence of the integral.
Now we can write the general solution to (\ref{appeqj})
\begin{eqnarray}
j_i(r)=j_P(r)+C_1 j_{H_1}(r)+C_2 j_{H_2}(r).
\end{eqnarray}
To determine $C_1$ and $C_2$ we need to find the asymptotic behavior of $j_i(r)$
\begin{eqnarray}
j_i^{(1)}(r)=C_1 r^4+r^3 \partial _v\beta _i+\frac{C_2-8 M C_1}{4}+O\left(\frac{1}{r}\right).
\end{eqnarray}
The coefficient $C_1$ multiplies a non-normalizable metric deformation, and so is forced to zero by our choice of boundary conditions. The other integration constant $C_2$ leads to a nonzero value for $T_{0i}$. We can remove this ambiguity by demanding that $u^\mu T^{(n)}_{\mu\nu}=0$, thus $C_2$ can be set to zero. In summary,
\begin{eqnarray}
j_i(r)&=&-r^4 f(r)\int _r^{\infty }dx x f(x) \zeta _i(x) \int _x^{\infty }\frac{dy}{y^5 f(y)^2} \nonumber\\
&& +r^4 f(r) \left(\int _r^{\infty }\frac{dx}{x^5 f(x)^2}\right) \left(r^3 \partial _v\beta _i+\int _r^{\infty }dx \left[x f(x) \zeta _i(x)+3 x^2 \partial _v\beta _i\right]\right). \label{appjsol}
\end{eqnarray}
Having the expression of $j_i(r)$, $a_i(r)$ is obtained by integrating (\ref{appeq1})
\begin{eqnarray}
a_i(r) = -g\int_r^\infty dr \frac{x^3}{\sqrt3 Q} \left[ \frac{x}{2} \left( \frac{j_i'(x)}{x^3} \right)'-S_{ri}(x)\right],
\end{eqnarray}
where we chose the gauge to make $a_i(r)$ vanishes at infinity.

Since we are interested in boundary physics, we present asymptotic behaviors of these functions here:
\begin{eqnarray}
j_i(r) &\approx& r^3 \partial _v\beta _i-\frac{2 \sqrt{3} Q}{5 r} \frac{e}{g}F_{v i}^{\text{\rm ext}}+\frac{1}{r^2} \left(-Q^2 \frac{j_i\left(r_+\right)}{r_+^4}-\frac{Q}{2 r_+} \left(\partial _iQ+Q \partial _v\beta _i\right)\right. \nonumber\\
&& \left.+\frac{r_+ Q}{2 \sqrt{3}} \frac{e}{g}F_{v i}^{\text{\rm ext}}\right), \\
a_i(r) &\approx& \frac{1}{r} eF_{v i}^{\text{\rm ext}}+\frac{1}{r^2} \left(\sqrt{3} gQ \frac{j_i\left(r_+\right)}{r_+^4}+\frac{\sqrt{3}g}{2 r_+} \left(\partial _iQ+Q \partial _v\beta _i\right)-\frac{r_+}{2} eF_{v i}^{\text{\rm ext}}\right).
\end{eqnarray}
Still we have undetermined constant, $j_i(r_+)$. Taking $r\to r_+$ limit to (\ref{appjsol}) yields an equation of $j_i(r_+)$. After some calculations we get
\begin{eqnarray}\label{j at rp}
\frac{j_i\left(r_+\right)}{r_+^4}&=&\frac{2 r_+^2 \left(2 r_+^4+r_+^2 r_-^2+r_-^4\right) \partial _v\beta _i-Q \left(\partial _iQ+Q \partial _v\beta _i\right)-\sqrt{3} r_+^2 Q \frac{e}{g}F_{v i}^{\text{\rm ext}}}{6 r_+ Q^2-2 r_+^3 \left(-2 r_+^4+r_+^2 r_-^2+r_-^4\right)} \nonumber\\
&=&\frac{2 \left(2 r_+^6+Q^2\right) \partial _v\beta _i-Q \left(\partial _iQ+Q \partial _v\beta _i\right)-\sqrt{3} r_+^2 Q \frac{e}{g}F_{v i}^{\text{\rm ext}}}{8 M r_+^3}.
\end{eqnarray}
From the definition
\begin{eqnarray}
j_i(r) &=& j_\beta(r) \partial_v \beta_i + j_Q(r) (\partial_i Q+Q\partial_v \beta_i) +j_F(r) F^{\rm ext}_{vi},
\end{eqnarray}
we can read off the result (\ref{jis}).

\section{Current expression and useful relations}\label{relations}

To obtain the current expression of (\ref{familiar current}), we massage (\ref{first order current}) and get following form
\begin{eqnarray}
J_{(1)}^{\mu }= \frac{1}{g} \left[-\frac{\sqrt{3} Q}{4 M} \left(2 r_+^3+\frac{Q^2}{r_+^3}\right) (3u^{\lambda } \partial _{\lambda }u^{\mu }  +  P^{\mu \lambda } \partial_\lambda \log Q  )         +\left(\frac{3 Q^2}{4 M r_+}+r_+\right) u^{\lambda } F_{\lambda  \mu }^{\text{ext}}\right]~~.
\end{eqnarray}
We need to change first term into derivatives of physical quantities. For this, we use more useful form of (\ref{conservation}),
 \begin{eqnarray}
u^{\lambda } \partial _{\lambda }u^{\mu }=\frac{\sqrt{3} Q}{4 M}\frac{e}{g} u^{\lambda } F_{\lambda  \mu }^{\text{ext}}-\frac{1}{4 M} P^{\mu  \nu } \partial _{\nu }M~.
 \end{eqnarray}
Using this equation and other relations between physical quantities, one can get our result (\ref{familiar current}). We write down the useful relations for this algebra as
\begin{eqnarray}
&&Q=\frac{\mu  r_+^2}{\sqrt{3} g},\\
&&r_+^2 r_-^2+r_-^4=\frac{\mu ^2 r_+^2}{3 g^2}=2 M-r_+^4,\\
&&T=\frac{1}{2 \pi  r_+^3} \left(2 r_+^4-r_+^2 r_-^2-r_-^4\right)=\frac{6 r_+^2-\mu ^2/g^2}{6 \pi  r_+}=\frac{3 r_+^4-2 M}{2 \pi  r_+^3}~.
\end{eqnarray}

\end{document}